%% file: ICVGIP-Latex-Template.tex
\documentclass[sigconf, review=false]{acmart}
\usepackage{booktabs}
\usepackage{siunitx}
\usepackage{multirow}
\usepackage{multirow}
\usepackage{subcaption}
\usepackage{booktabs}
\usepackage{amsmath}
\usepackage{graphicx}
\usepackage{graphicx}
\usepackage{float}
\usepackage{algpseudocode}
\usepackage[ruled,vlined]{algorithm2e}
\SetCommentSty{mycommfont}
\usepackage{lineno,xcolor}
\SetKwInput{KwInput}{Input}                
\SetKwInput{KwOutput}{Output}
\graphicspath{ {./images/} }
\usepackage{pythonhighlight}
\usepackage{tabularx}
\usepackage{orcidlink}

\usepackage{hologo} 
\usepackage{booktabs} 
\usepackage{threeparttable}

\setcopyright{rightsretained}

\acmConference[Preprint]{Preprint}{August}{2024}
\acmYear{2024}
\copyrightyear{2024}

\acmPrice{15.00}

\begin{document}
\title{Integrating Edge Information into Ground Truth for the Segmentation of the Optic Disc and Cup from Fundus Images}
\author{Yoga Sri Varshan V}
\orcid{0009-0007-8440-8060}
\affiliation{
 \institution{Indian Institute of Information Technology, Design and Manufacturing, Kancheepuram}
\streetaddress{Vandalur-Kelambakkam Road}
\city{Chennai}
 \state{Tamil Nadu}
 \country{India}
 \postcode{600127}  
}

\author{Hitesh Gupta Kattamuri}
\orcid{0009-0000-2751-1802}
\affiliation{
 \institution{Indian Institute of Information Technology, Design and Manufacturing, Kancheepuram}
\streetaddress{Vandalur-Kelambakkam Road}
\city{Chennai}
 \state{Tamil Nadu}
 \country{India}
 \postcode{600127}  
}

\author{Subin Sahayam}
\orcid{0000-0003-1129-895X}
\affiliation{
 \institution{Indian Institute of Information Technology, Design and Manufacturing, Kancheepuram}
\streetaddress{Vandalur-Kelambakkam Road}
\city{Chennai}
 \state{Tamil Nadu}
 \country{India}
 \postcode{600127}  
}

\author{Umarani Jayaraman}
\orcid{0000-0002-9676-6291}
\affiliation{
 \institution{Indian Institute of Information Technology, Design and Manufacturing, Kancheepuram}
\streetaddress{Vandalur-Kelambakkam Road}
\city{Chennai}
 \state{Tamil Nadu}
 \country{India}
 \postcode{600127}  
}

\renewcommand{\shortauthors}{}

\begin{abstract}
Optic disc and cup segmentation helps in the diagnosis of glaucoma, myocardial infarction, and diabetic retinopathy. Most deep learning methods developed to perform segmentation tasks are built on top of a U-Net-based model architecture. Nevertheless, U-Net and its variants have a tendency to over-segment/ under-segment the required regions of interest. Since the most important outcome of the segmentation is the value of cup-to-disc ratio and not the segmented regions themselves, we are more concerned about the boundaries rather than the regions under the boundaries. This makes learning edges important as compared to learning the regions. In the proposed work, the authors aim to extract both edges of the optic disc and cup from the ground truth using a Laplacian filter. Next, edges are reconstructed to obtain an edge ground truth in addition to the optic disc-cup ground truth. Utilizing both ground truths, the authors study several U-Net and its variant architectures with and without optic disc and cup edges as a target, along with the optic disc-cup ground truth for segmentation. The authors have used the REFUGE benchmark dataset and the Drishti-GS dataset to perform the study, and the results are tabulated for the dice and the Hausdorff distance metrics. In the case of the REFUGE dataset, the mean dice score for the optic disc has improved from 0.7425 to 0.8859 while the mean Hausdorff distance metric has reduced from 6.5810 to 3.0540 for the baseline U-Net model. Similarly, the mean dice score for the optic cup has improved from 0.6970 to 0.8639 while the mean Hausdorff distance metric has reduced from 5.2340 to 2.6323 for the same model. Similar performance improvement has been observed in the Drishti-GS dataset as well. Compared to the baseline U-Net and its variants (i.e) the Attention U-Net and the U-Net++, the models that learn integrated edges along with the optic disc and cup regions performed well in both validation and testing datasets. 
\end{abstract}

%
%

\ccsdesc[500]{Computer systems organization~Embedded systems}
\ccsdesc[300]{Computer systems organization~Redundancy}
\ccsdesc{Computer systems organization~Robotics}
\ccsdesc[100]{Networks~Network reliability}

\keywords{Optic Disc, Optic Cup, Edge Integration, Segmentation, Cup-To-Disc Ratio, Retinal Images, Glaucoma, REFUGE dataset, Drishti-GS dataset}

\maketitle

\input{samplebody-conf}

\bibliographystyle{ACM-Reference-Format}
\bibliography{ICVGIP-Latex-Template}

\appendix

\end{document}

%% file: samplebody-conf.tex
\section{INTRODUCTION}

\begin{figure}[htb]
    \centering
        \subfloat[\centering Normal eye vs Glaucoma eye (enlarged optic cup)]
        {\includegraphics[width=0.525\linewidth]{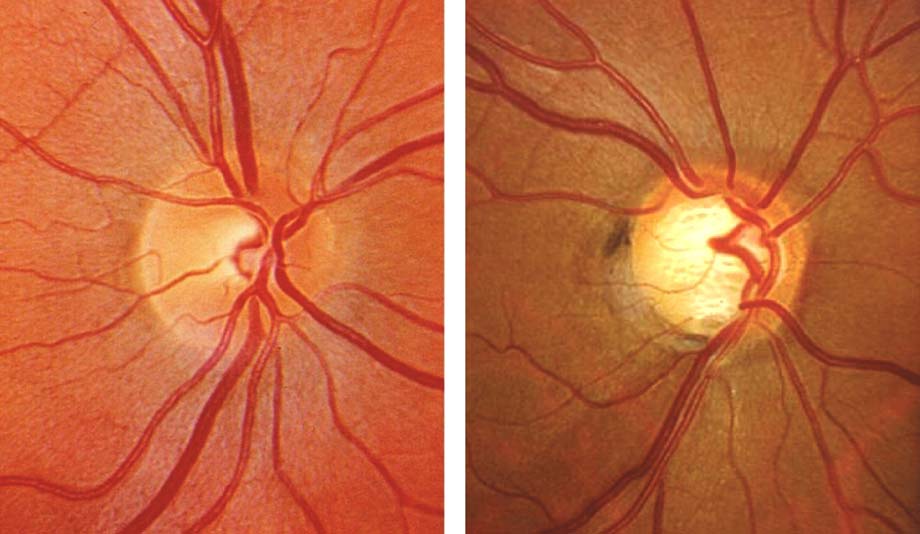}}
        \quad
        \subfloat[\centering Optic disc and cup]
        {\includegraphics[width=0.29\linewidth,]{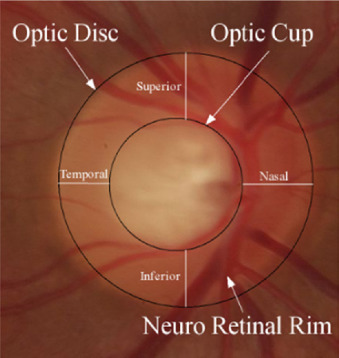}}
        \caption{Retinal fundus image}
        \Description{Image showing the optic disc and cup}
    \label{fig:fundus}
\end{figure}

Every year, millions of people around the world are affected by eye disorders such as glaucoma, retinopathy, cardiac infection, etc. \cite{WHO2023} For instance, glaucoma, a chronic eye condition that harms the optic nerve, is second only to cataract in terms of causes of blindness \cite{Quigley262}. An example of a normal eye and a glaucoma-affected eye is shown in Figure \ref{fig:fundus}(a). 
For instance, with almost 60 million cases recorded worldwide in 2010 \cite{Quigley262}, glaucoma ranked second after cataract as the main cause of vision loss in individuals. In 2020, this disease was anticipated to afflict more than 80 million people\cite{Quigley262}. 
These eye illnesses require the fundus image to be segmented clearly into vessels, the optic disc, the optic cup, and the background. In particular, the optic disc-to-cup ratio \cite{garway1998vertical} is the most useful technique for glaucoma diagnosis. A sample eye with optic disc and cup is shown in Figure \ref{fig:fundus}(b).


Segmentation of the optic disc and cup of a fundus image is a computer vision task. Difficulties in this field include handling image uncertainties that cannot be otherwise eliminated, including various sorts of information that is incomplete, noisy, imprecise, and fragmentary \cite{Hussien_2021}. Accurately predicting the boundary between adjacent segments is a difficult task. 


The most prominent dataset providing high-resolution retinal fundus images and their optic disc-cup ground truths is the REFUGE benchmark dataset \cite{orlando2020refuge} which was organized as a half day Challenge in conjunction with the 5th MICCAI Workshop on Ophthalmic Medical Image Analysis (OMIA), a Satellite Event of the MICCAI 2018 conference in Granada, Spain in 2020. The goal of the challenge was to evaluate and compare automated algorithms for glaucoma detection and optic disc/cup segmentation on a common dataset of retinal fundus images. The dataset contains 1200 images with optic disc and cup segmented ground truth. It is one of the largest datasets, and its data and evaluation remains open to encourage further developments and ensure a proper and fair comparison of those new proposals. Another dataset which is also used for optic disc and cup segmentation is the Drishti-GS dataset \cite{sivaswamy2014drishti}, developed by IIIT, Hyderabad, India in collabration with Aravind Eye Hospital, Madurai, India. It has 101 images with optic disc and cup segmented ground truth.

Various deep learning based approaches have given excellent results and there is a lot of active research presently going on in this field. Some popular semantic segmentation models for optic disc and cup segmentation are U-Net \cite{ronneberger2015u}, V-Net \cite{milletari2016v}, Attention U-Net \cite{oktay2018attention} and Swin U-Net \cite{cao2023swin}. Although, the majority of proposed methods give fairly good results, they focus more on the model architecture rather than the input and segmented data images. New approaches using powerful and robust deep learning models such as transformers are also constantly being developed in this field of research.

\subsection{Need for integrating edges}
The final objective of segmenting the optic disc and cup from a retinal fundus image is to calculate the optic cup-to-disc ratio. In the paper \cite{tadisetty2023identifying}, Tadisetty et al propose a method to segment using the popular U-Net model and then apply canny edge detection and dilation to post-process the segmented output to get the edges of the optic disc and cup. Most approaches employ a similar workflow, feeding large amounts of retinal fundus images and their segmented ground truths for training the model. This results in the models performing well on the dice score values, but often over-segmenting or under-segmenting the input image which is evident from the high  Hausdorff distances from the results. A similar problem was encountered by Sahayam et al for the BraTS20 3D dataset and was solved by learning edges as targets along with the brain tumor regions in the paper \cite{sahayam2024integrating}. Hence, there is a need to learn the optic disc and cup edges along with the disc-cup regions in order to better segment fundus images. 

\subsection{Contribution of the work}
While there are existing methods to segment optic disc-cup regions in fundus images, the authors suggest that any deep learning model will learn better if edges are integrated along with the regions of interest. The authors aim to solve this problem by stacking the optic disc and cup edges in the ground truths. The contributions of the paper are given below,

\begin{enumerate}
    \item Both the optic disc and cup edges are extracted from the ground truth using a 2D Laplacian filter and stacked as a part of the ground truth.
    \item The edges and the disc-cup regions are used as targets for deep learning architectures. 
    \item The proposed method has been evaluated on the extensive REFUGE dataset containing 1200 retinal fundus images with their ground truths and the Drishti-GS dataset containing 101 retinal fundus images with their ground truths.
    \item The Hausdorff distance is used as a metric to evaluate the correctness of the predictions.
    \item The dice and Hausdorff distance scores have improved when the edges are integrated with the disc-cup regions.
\end{enumerate}

The rest of the paper is organized as follows. In Section \ref{RW}, related work has been discussed. The proposed methodology is given in Section \ref{PM}. In Section \ref{ER}, experimental results have been discussed. Conclusions and future work are discussed in Section \ref{con}. References are given in the last section.

\section{RELATED WORK}
\label{RW}

Recent developments in segmenting optic disc and cup are driven by computer-aided methodologies for applications in clinical ophthalmology.

\subsection{U-Net based architectures}
Most of the deep learning architectures proposed for optic disc and cup segmentation are based on U-Net \cite{ronneberger2015u}. The authors of \cite{yu2019robust}, propose a modified U-Net architecture, which combines the widely adopted pre-trained ResNet-34 \cite{koonce2021resnet} model as encoding layers with classical U-Net decoding layers. The model achieved an impressive mean optic disc dice value of 0.974 and mean optic cup dice value of 0.888 on the Drishti-GS data set. In the paper \cite{10322515}, Hanifa Suwandoko et al have developed an interesting approach by performing optic disc and cup localization as the pre-processing step, followed by training the dataset on the U-Net model, followed by an ellipse fitting algorithm as the post-processing step. This strategy has been found to achieve an optic disc segmentation mean dice score of 0.979 in the training set of Drishti-GS and 0.942 in the REFUGE dataset. The optic cup segmentation has been found to achieve a mean dice score of 0.948 for the Drishti-GS training set and a dice score of 0.843 for the REFUGE dataset.

The MICCAI workshop on Ophthalmic Medical Image Analysis \cite{fu2020ophthalmic} produced two papers that are also based on the U-Net baseline. In one of them \cite{kamble2020optic}, Ravi Kamble et al introduce an approach incorporating optic disc, optic cup, and optic fovea analysis together. The EfficientNet-B4 \cite{tan2019efficientnet} model serves as the foundation for a unique method for the detection of an optic disc with a cup and fovea that is presented in this study. In U-Net++ \cite{zhou2018unet++}, the retrieved features from the EfficientNet are used with skip connections to segment data precisely. The proposed method achieved a mean dice score of 0.957 for optic disc segmentation on the REFUGE dataset. The second paper \cite{liu2020multi} proposes a novel optic disc segmentation network by developing two separate modules, namely an atrous convolution spatial pyramid pooling module and a light U-Net module. The authors first use ResNet-101 \cite{wu2019wider} as a basic network to extract hierarchical characteristics. Atrous convolution and spatial pyramid pooling module is then used to incorporate global spatial information in high-level semantic features. Finally, they integrate the spatial information by feature fusion to get the segmentation results. All the papers have worked on industry standard datasets such as the REFUGE and the Drishti-GS datasets.

\subsection{RCNN based architectures}
One prominent work involving RCNN's for this task is the architecture developed by Haidar Almubarak et al \cite{almubarak2020two}. When compared to complex techniques that are considered the state-of-the-art in the literature, their approach is based on a straightforward two-stage Mask-RCNN. They locate and crop around the optic nerve head in the first stage, then provide the cropped image as input for the second stage. To create the final segmentation, the second stage network is trained using a weighted loss. By combining the cropping output from the first stage with the original training image, they suggest a novel fine-tuning technique to further enhance the detection in the first stage. This new detection network is trained using multiple scales for the region proposal network anchors. The authors claim to have achieved a mean dice score of 0.854 and 0.947 for the optic cup and disc respectively on the REFUGE dataset.

\subsection{Transformer based architectures}
Recently, transformers are being increasingly used in the field of computer vision. They seem to learn image features that incorporate large context while keeping high spatial resolutions. The authors of \cite{li2021medical}, propose a novel squeeze-and-expansion transformer Segtran, the core of which is a squeezed attention block that regularizes the self-attention of transformers, and an expansion block that learns diversified representations. The Segtran model with EfficientNet-B4 \cite{tan2019efficientnet} achieved a dice score of 0.872 for the optic cup and 0.961 for the optic disc, on the REFUGE dataset.

\section{PROPOSED METHODOLOGY}
\label{PM}
\begin{figure*}[htb]
    \includegraphics[width=0.65\linewidth]{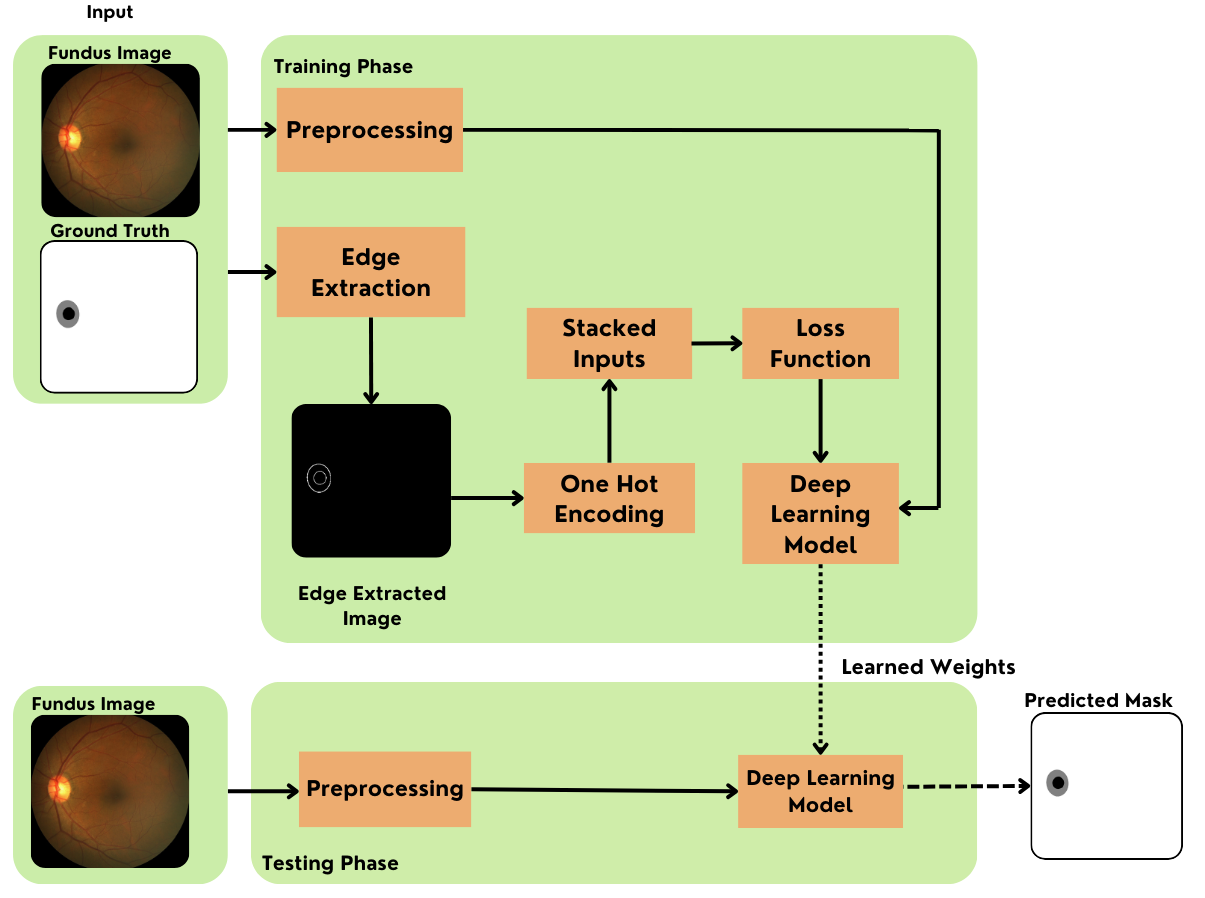}
    \caption{Proposed Workflow}
    \label{fig:proposed_workflow}
\end{figure*}

\begin{figure}[htb]
    \centering
        \subfloat[\centering Ground truth]
        {\includegraphics[width=0.2\linewidth]{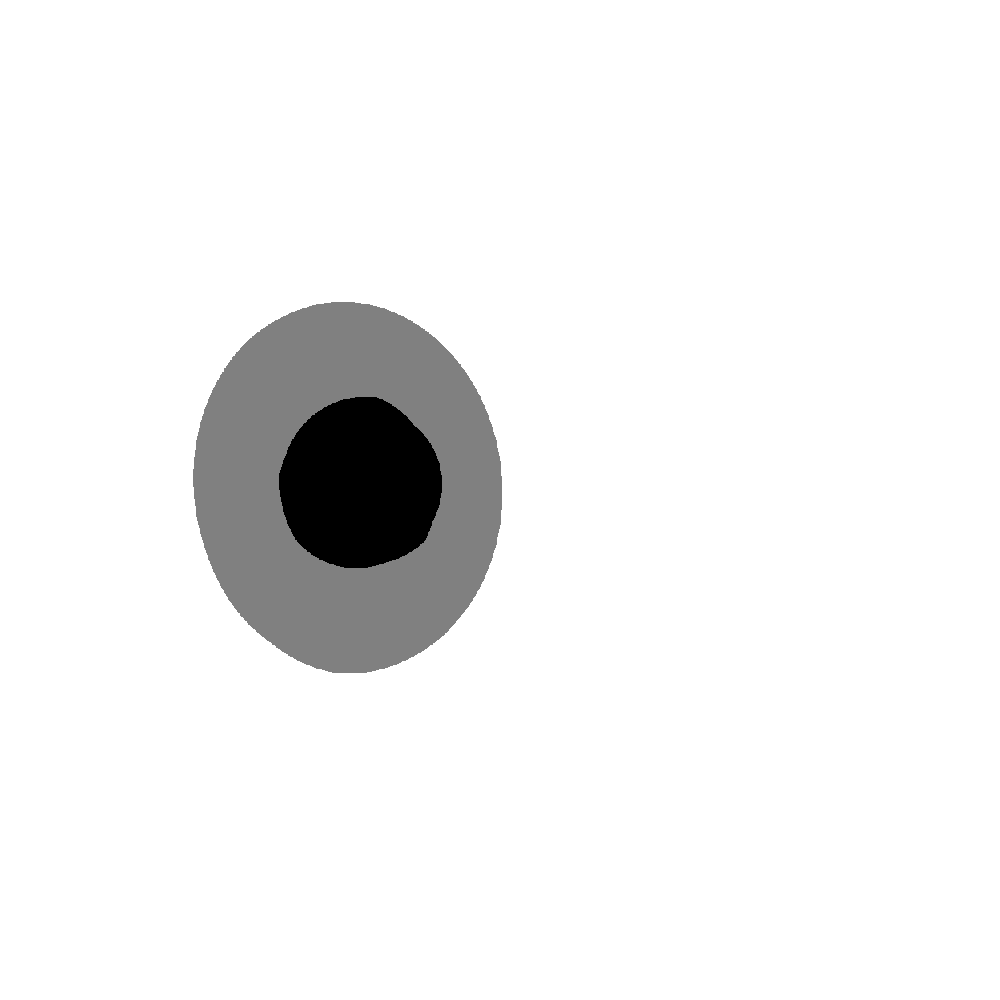}}
        \qquad
        \subfloat[\centering Extracted edges]
        {\includegraphics[width=0.2\linewidth]{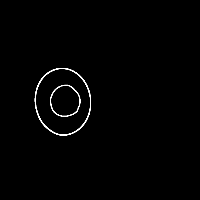}}
        \caption{Edge ground truth generated using the convolution operation on the ground truth}
    \label{fig:edge_mask}
\end{figure}

The authors have proposed a three-step process for segmenting optic disc-cup regions. Initially, it includes extracting edges from the ground truth ($x$ x $y$ x $1$) and re-coding them into categories (one-hot - ($x$ x $y$ x $3$)). The re-coded one-hot edge ground truth ($x$ x $y$ x $3$) is stacked with the original ground truth (re-coded to three categories - ($x$ x $y$ x $3$)) resulting in ($x$ x $y$ x $5$). The extra background channel is dropped. This newly created multi-channel ground-truth ground truth has been used for training the deep learning models with a custom loss function. The authors have performed these experiments on the REFUGE dataset \cite{orlando2020refuge} and the Drishti-GS dataset \cite{sivaswamy2014drishti}. An example of a ground truth image and its edge extracted image is shown in Figure \ref{fig:edge_mask}, and its corresponding one hot encoded images are shown in Figure \ref{fig: mul}. The overall workflow is shown in Figure \ref{fig:proposed_workflow}.

\subsection{Extracting Optic Disc and Cup Edges}
The optic disc and cup are critical anatomical features in the retina, whose dimensions and relative sizes are important indicators of conditions such as glaucoma. The edges of these regions are often not given much importance as the regions themselves. Integrating edges into the ground truth for training is a great way for the model to understand the underlying boundaries. In the paper \cite{sahayam2024integrating}, an edge extraction procedure using a 3D Laplacian operator for brain tumor regions is employed. The authors used a similar technique, using a 2D Laplacian operator, to extract the edges of the optic disc and the optic cup, separately. A Laplacian operator will enhance any feature with a sharp discontinuity \cite{wang2007laplacian}. It is a simple edge extraction filter that is sensitive to noise \cite{wang2007laplacian}. Since the datasets' ground truths are noise-free, the 2D Laplacian operators are ideal for edge extraction. The Laplacian filter used to convolve over the ground truth image to obtain the edge ground truth is given below,

\[
\begin{bmatrix}
-1 & -1 & -1 \\
-1 & 8 & -1 \\
-1 & -1 & -1
\end{bmatrix}\]


The edge detection process is applied separately to the disc and cup regions to highlight their boundaries. The convolution with the Laplacian filter highlights the edges by emphasizing areas of high-intensity change, which correspond to the boundaries between the optic disc/cup and the surrounding retina. The filter highlights these areas by enhancing the contrast between pixels at the boundaries of different regions. The entire process consists of three steps,

\begin{enumerate}
\item Ground truth images are re-coded such that pixels belonging to the optic disc are marked as 1, and those belonging to the optic cup as 2. The background and other areas are treated differently (marked as 0) as shown in Figure \ref{fig: mul}(a).

\item The original ground truth is first converted into a one hot ground truth image where pixels belonging to the disc are set to 1, and all other pixels are set to 0, as shown in Figure \ref{fig: mul}(b). The convolution operation is then applied to this one-hot ground truth. The resulting output highlights the edges of the disc. After convolution, any non-zero values (indicating edge presence) are set to 1 to create a clear one-hot optic disc edge map as in Figure \ref{fig: mul}(e).

\item In the next step, a similar process is followed for the cup edges. However, the one hot ground truth image is created for the optic cup in this scenario. The convolution highlights the edges of the cup, and the output is binarized to create an optic cup edge map as illustrated in Figures \ref{fig: mul}(c) and \ref{fig: mul}(f).
\end{enumerate}

\begin{algorithm}[hbt]
\DontPrintSemicolon
  \KwInput{Zero Padded 2D Image $I_{in}$, 2D Laplacian filter $\mathcal{F}$.}
  \KwOutput{2D Edge Extracted Image $I_{out}$}
  \For{ $\forall (x, y)  \in I_{in}$}{
    \For{$\forall (i, j)  \in \mathcal{F}$}{
        \tcc{Convolution Operation of Filter $\mathcal{F}$ over Input $I_{in}$}
        \tcc{$a, b$ are the half filter Size of Filter $\mathcal{F}$}
        $T(x,y) \gets \sum_{i=-a}^{a}\sum_{j=-b}^{b} F(i,j) * I_{in}(x+i,y+j)$\\
    }
  }
  \tcc{Initialize $I_{out}$ of dimension $(x, y)$ to $0$}
  $I_{out} \gets 0$\\
  \For{ $\forall (x, y)  \in I_{in}$}{
  \tcc{For all non-zero values in $T(x,y)$, replace the 0 in $I_{out}(x, y)$ with $I_{in}(x, y)$ }
    \If{$T(x,y) \ne 0$}{
            $I_{out}(x, y) \gets I_{in}(x, y)$ \\
        }
    }
    \Return{$I_{out}$}
\caption{Extraction of Edges from 2D Images}
\label{alg:edge2d}
\end{algorithm}
The process of edge extraction from the ground truth image is mentioned in Algorithm \ref{alg:edge2d}.

\subsection{Stacking for a Multi-Channel Mask}
\begin{figure}[hbt]
    \centering
    \captionsetup[subfigure]{justification=centering}
    
    \subfloat[Ground truth]{{\includegraphics[width=0.2\linewidth]{g0008_mask_c.png} }}
    \quad
    \subfloat[Optic disc region]{{\includegraphics[width=0.2\linewidth]{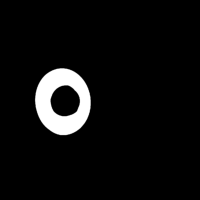} }}
    \quad
    \subfloat[Optic cup region]{{\includegraphics[width=0.2\linewidth]{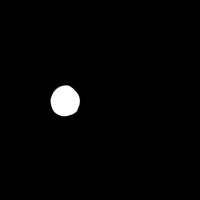} }}

    \subfloat[Background region]{{\includegraphics[width=0.2\linewidth]{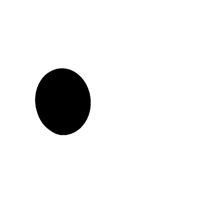} }}
    \quad
    \subfloat[Optic cup and optic disc edges]{{\includegraphics[width=0.2\linewidth]{g0008_mask_od_edge_c_new.png} }}
    \quad
    \subfloat[Optic cup edge]{{\includegraphics[width=0.2\linewidth]{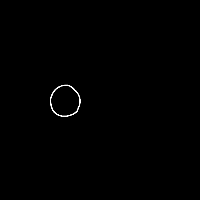} }}
    
    \caption{Sample set of a ground truth image and its corresponding one hot images of regions and edges}
    \label{fig: mul}
\end{figure}

The authors have employed a stacking process that involves combining multiple one-hot images into a single multi-channel image. Stacking, in the context of image processing and machine learning, refers to the process of aligning and combining several two-dimensional arrays (images or masks) along a new dimension to form a multi-dimensional array. After obtaining the one-hot image for the optic disc, the optic cup, and their respective edges, the stacking process combines these into a single multi-channel ground truth. Each channel represents a different aspect of the segmentation task as shown in Figure \ref{fig: mul}. Each image in stack is as follows,

\begin{enumerate}
    \item Optic cup ground truth: A binary ground truth where pixels belonging to the optic cup are marked as 1, and all others are 0.
    \item Optic disc ground truth: A binary ground truth where pixels belonging to the optic disc are marked as 1, and all others are 0.
    \item Disc edge ground truth: A binary ground truth highlighting the edges of the optic disc.
    \item Cup edge ground truth: A binary ground truth highlighting the edges of the optic cup.
    \item Background: A channel for the background is also included.
\end{enumerate}

By including both the segmentation masks and the edge maps as separate channels, the model receives detailed information about the regions of interest (optic disc and cup) and their boundaries. This can lead to more accurate segmentation, as the model can leverage the explicit boundary information provided by the edge maps.

\subsection{Training with a Custom Loss Function}
The above two steps are followed by training the deep learning model on the input image and their modified ground truth. The authors have proposed a custom loss function based on focal loss for the training process. The mathematical equation for focal loss is as follows,

\begin{equation}
   \textrm{Focal Loss $L(y,\hat{y})$ = }-\alpha y(1-\hat{y})^{\gamma}log(\hat{y})-(1-y)\hat{y}^{\gamma}log(1-\hat{y})
\end{equation}

The $\alpha$ parameter controls the weight. A modulating factor called $\gamma$ makes hard examples more important by assigning them more weight and simple ones less important. In the proposed study, all the deep learning models are trained using a $\gamma$ value of $2$.

The value of alpha proposed by the authors varies depending on the target class as follows,
 where \[
    \alpha= 
\begin{cases}
0.9,& \text{if }  \textrm{target} \in \{\textrm{Optic cup edge}\}\\
   0.8,& \text{if } \textrm{target} \in \{\textrm{Optic disc edge}\}\\
    0.1,              & \textrm{Background}
\end{cases}
\]

The optic cup is given the most weightage since it is the most difficult region to learn. The background is given the least weightage, since more than 80\% of the ground truth (conservative estimate) is of the background. 

\section{EXPERIMENTAL AND RESULTS}
\label{ER}
\subsection{Dataset}
\begin{figure}[htb]
    \centering
        \subfloat[\centering Fundus Image]
        {\includegraphics[width=0.25\linewidth]{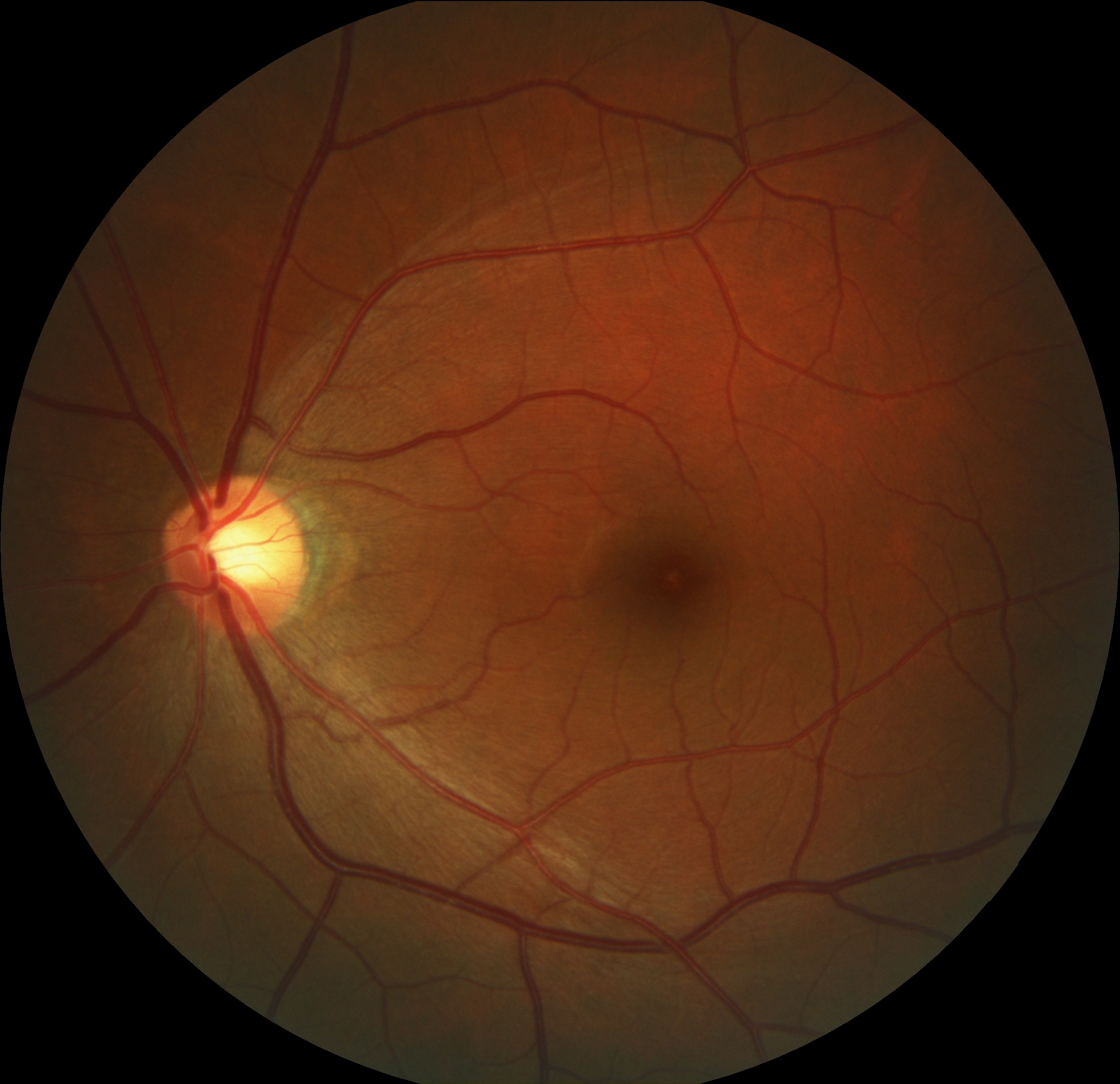}}
        \quad
        \subfloat[\centering Ground Truth]
        {\includegraphics[width=0.25\linewidth]{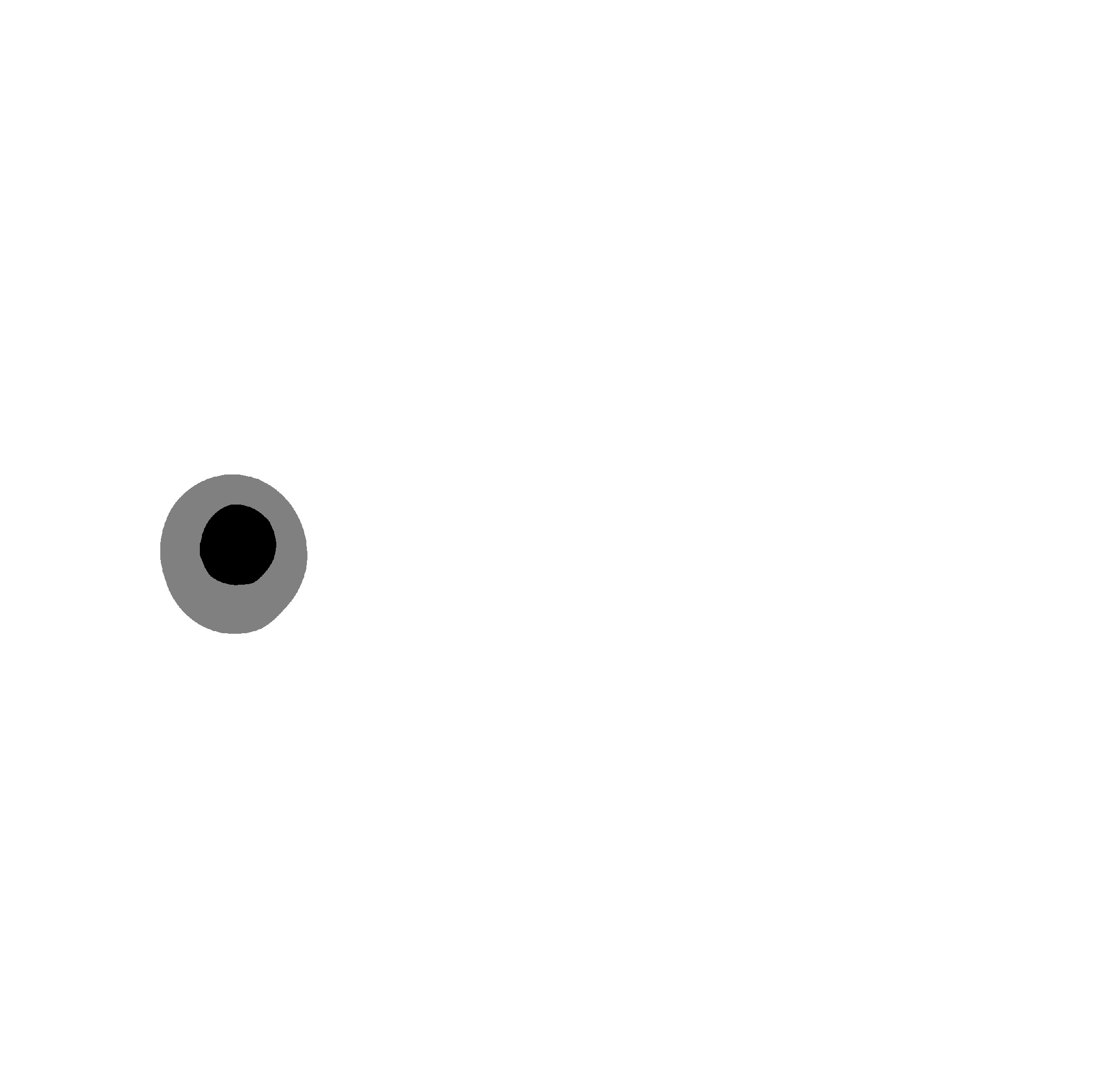}}
        \caption{Sample fundus image and the corresponding ground truth from REFUGE}
    \label{fig: samples}
\end{figure}

The proposed algorithm has been evaluated on the REFUGE dataset \cite{orlando2020refuge} and the Drishti-GS dataset \cite{sivaswamy2014drishti}. The REFUGE dataset was organized as a challenge as a part of MICCAI 2020 conference in Lima, Peru. The dataset consists of 1200 images of fundus each of size $2124$ x $2056$ pixels. The optic disc and cup segmentation masks of all the images were provided. The authors have chosen the mentioned dataset as it is one of the benchmarks for optic disc-cup segmentation. In Figure \ref{fig: samples}, a sample of an image from the dataset and its corresponding ground truth is shown. The Drishti-GS dataset has 101 images, each of size $2049$ x $1751$ pixels. The optic disc and cup segmentation masks of all the images were provided here as well.

\subsection{Implementation Details}
The authors have used various popular deep learning-based models for the study, namely U-Net \cite{ronneberger2015u}, Attention U-Net \cite{oktay2018attention} and U-Net++ \cite{zhou2018unet++}. The models have been trained and tested on ground truth masks with and without optic disc and optic cup edge targets for 50 epochs with a batch size of 32. In the case of the REFUGE dataset, all the three models were used for the experiments and in the case of the Drishti-GS dataset, only U-Net and Attention U-Net were used. Adam \cite{kingma2014adam} has been used as an optimizer with an initial learning rate $\alpha$ of $0.01$. To ensure a fair comparison, all models employ ReLU \cite{agarap2018deep} as their activation functions. In the last layer, softmax activation was used to predict the final optic disc/cup region and/or edge.

The workflow utilizes a custom loss function based on focal loss, adapts to class imbalance, and employs data generators for efficient memory usage. The stacked one-hot representation would measure $2124$ x $2056$ x $3$ if the training comprised solely the optic disc and cup regions in the case of REFUGE dataset images. It would measure $2049$ x $1751$ x $3$ in the case of Drishti-GS dataset images. The final dimension would represent the background. Alternatively, the dimensions would be $2124$ x $2056$ x $5$ and $2049$ x $1751$ x $5$ for REFUGE and Drishti-GS dataset images respectively if the training used both optic disc and optic cup edges as targets. The authors have employed a training, validation, and testing split of 0.7, 0.1, and 0.2. Additionally, the authors have also performed a 5-fold cross validation experiment on the U-Net model trained with and without edge targets. The above mentioned experiments were performed for the both the datasets separately. All the experiments were conducted on an NVIDIA A100 GPU with 40GB RAM. 

\subsection{Evaluation Metrics}
For measuring the correctness of the proposed segmentation, the authors have made use of the dice score and the bi-direction Hausdorff distance.

\subsubsection{Dice Score}
The dice score is a measure used to evaluate the accuracy of a model's predictions. It is particularly useful in the context of binary classification problems. The formula for calculating the dice score is defined in equation \ref{dice}.

\begin{equation}
    \textrm{Dice score =} \frac{2*TP}{(TP + FP) + (TP + FN)}
    \label{dice}
\end{equation}

This equation essentially calculates the proportion of correct positive predictions (true positives) over the average size of both the predicted positives and actual positives. The dice score is a way to capture the model's accuracy by considering both the precision (the proportion of true positives among all positive predictions) and recall (the proportion of true positives detected among all actual positives). Here true positive (TP) is the total number of true class instances correctly predicted as true, false positive (FP) is the total number of false class instances incorrectly predicted as true, and false negative (FN) is the total number of true class instances incorrectly predicted as false.

\subsubsection{Hausdorff Distance}
In the domain of image segmentation, the Hausdorff distance is a significant metric for assessing the dissimilarity between two sets of points, which in this context, usually represent the edges of segmentation masks. It calculates the greatest of all the distances from a point in one set to the closest point in the other set. This metric is crucial for determining how closely a predicted segmentation ground truth aligns with the ground truth, especially around the edges. A higher Hausdorff distance suggests a larger discrepancy between the predicted and actual segmentation, indicating poorer model performance.

The Hausdorff distance can be computed in a one-sided manner from set X to set Y, and vice versa, using the formulas given in equation \ref{eq4} and \ref{eq5}.
\begin{equation}
        \textrm{$\hat{H}$(X,Y) = } max(min_{x \in X}d(x,Y))
        \label{eq4}
\end{equation}
\begin{equation}
    \textrm{$\hat{H}$(Y,X) = } max(min_{y \in Y}d(X,y))\
    \label{eq5}
\end{equation}
However, this one-sided Hausdorff distance does not satisfy the symmetry property required of a metric, meaning it does not provide a consistent measure of distance regardless of the order of the sets. To address this, the bidirectional Hausdorff distance is used, defined in equation \ref{eq6}.
\begin{equation}
    \textrm{H(X,Y) = }max(\hat{H}(X,Y), \hat{H}(X,Y))
    \label{eq6}
\end{equation}
This bidirectional measure is considered a true metric and provides a more reliable assessment of similarity between two sets.

\subsection{Results}
The mean dice score and Hausdorff distance results of all the testing images with and without optic disc and cup edge integration for the optic disc and optic cup are tabulated in Table \ref{tab:mean_opticdisc}. The table contains the results for each of the models used in the experiments. For table \ref{tab:mean_opticdisc}, U-Net Edge, Attention U-Net Edge, and U-Net++ Edge correspond to the models trained on the edge-integrated ground truth images. All the models have mostly shown improved performances in both the metrics. 

In the case of the REFUGE dataset, the Attention U-Net Edge and the U-Net++ Edge models did not improve much the Hausdorff distance for the optic disc region. Nevertheless, there is very good improvement for both the optic disc and cup regions as far as the baseline U-Net model is concerned. In general, the authors also recognize the fact that Attention U-Net models focus more on relevant features for segmentation and less on background or irrelevant areas by introducing an attention mechanism to the U-Net architecture, specifically targeting the skip connections. This may have the reason for lower (or in some cases slight negative) improvements in the results concerning Attention U-Net Edge. The U-Net++ Edge model, however, performs significantly better than its counterpart, the U-Net++ model, which further emphasizes the effect of learning edge targets.

In the case of the Drishti-GS dataset, both U-Net Edge and Attention U-Net Edge models have shown decent improvement in the dice score metrics. However, the Attention U-Net Edge model did not improve much the Hausdorff distance for the optic disc and cup regions. The authors note the fact that even though the Drishti-GS dataset is a much smaller collection of fundus images than the REFUGE dataset, the consistent increase in performance significantly underscores the effect of learning edge targets along with the original ground truth.
\begin{table*}
\caption{Comparison results of mean dice score and Hausdorff distance}
\label{tab:mean_opticdisc}
\begin{tabularx}{\textwidth}{|l|l|X|X|X|X|}
\hline
 &  & \multicolumn{2}{c|}{Optic Disc} & \multicolumn{2}{c|}{Optic Cup} \\
\hline
Dataset & Model & Dice Score$\uparrow$ & Hausdorff $\downarrow$ & Dice Score$\uparrow$ & Hausdorff$\downarrow$ \\
\hline
\multirow{6}{*}{REFUGE} & U-Net \cite{ronneberger2015u} & 0.7425 & 6.581 & 0.6970 & 5.2340 \\
 & U-Net Edge & \textbf{0.8859} & \textbf{3.054} & \textbf{0.8639} & \textbf{2.6323} \\
\cline{2-6}
 & Attention U-Net \cite{oktay2018attention} & 0.8780 & \textbf{3.3394} & \textbf{0.8679} & 5.8275 \\
 & Attention U-Net Edge & \textbf{0.8982} & 4.692 & 0.8656 & \textbf{2.6262} \\
\cline{2-6}
 & U-Net++ \cite{zhou2018unet++} & 0.8166 & 4.3426 & 0.6538 & 4.5705 \\
 & U-Net++ Edge & \textbf{0.8811} & \textbf{4.3326} & \textbf{0.8505} & \textbf{4.0500} \\
\hline
\multirow{4}{*}{Drishti-GS} & U-Net \cite{ronneberger2015u} & 0.9268 & 27.7258 & 0.8990 & 4.8156 \\
 & U-Net Edge & \textbf{0.9464} & \textbf{12.2003} & \textbf{0.9102} & \textbf{4.4547} \\
\cline{2-6}
 & Attention U-Net \cite{oktay2018attention} & 0.9474 & \textbf{25.7457} & 0.8689 & \textbf{5.2318} \\
 & Attention U-Net Edge & \textbf{0.9493} & 45.7173 & \textbf{0.8837} & 7.2621 \\
\hline
\end{tabularx}
\end{table*}
\begin{table*}
\caption{Comparison results of median dice score and Hausdorff distance}
\label{tab:median_opticdisc}
\begin{tabularx}{\textwidth}{|l|l|X|X|X|X|}
\hline
 &  & \multicolumn{2}{c|}{Optic Disc} & \multicolumn{2}{c|}{Optic Cup} \\
\hline
Dataset & Model & Dice Score$\uparrow$ & Hausdorff $\downarrow$ & Dice Score$\uparrow$ & Hausdorff$\downarrow$ \\
\hline
\multirow{6}{*}{REFUGE} & U-Net \cite{ronneberger2015u} & 0.7618 & 5.8309 & 0.7042 & 5.0990 \\
 & U-Net Edge & \textbf{0.9014} & \textbf{2.8262} & \textbf{0.8777} & \textbf{2.2361} \\
\cline{2-6}
 & Attention U-Net \cite{oktay2018attention} & 0.9029 & 2.8284 & \textbf{0.8840} & 2.2360 \\
 & Attention U-Net Edge & \textbf{0.9165} & \textbf{2.6105} & 0.8826 & \textbf{2.0515} \\
\cline{2-6}
 & U-Net++ \cite{zhou2018unet++} & 0.8452 & 4.3426 & 0.7048 & 4.0000 \\
 & U-Net++ Edge & \textbf{0.9037} & \textbf{2.2174} & \textbf{0.8639} & \textbf{2.4237} \\
\hline
\multirow{4}{*}{Drishti-GS} & U-Net \cite{ronneberger2015u} & 0.9294 & 4.4721 & \textbf{0.9204} & \textbf{4.0000} \\
 & U-Net Edge & \textbf{0.9648} & \textbf{2.8284} & 0.9086 & 5.0000 \\
\cline{2-6}
 & Attention U-Net \cite{oktay2018attention} & 0.9473 & \textbf{4.1231} & 0.8576 & \textbf{5.8309} \\
 & Attention U-Net Edge & \textbf{0.9606} & 4.1621 & \textbf{0.8881} & 8.0622 \\
\hline
\end{tabularx}
\end{table*}
\begin{table*}[htbp]
\caption{Comparison results of cross validation dice scores for optic disc}
\label{tab:cross_validation}
\begin{tabular}{|l|l|*{6}{p{1.25cm}|}p{1.5cm}|}
\hline
 Dataset & Model & Fold 1 & Fold 2 & Fold 3 & Fold 4 & Fold 5 & Average & Median \\
\hline
\multirow{2}{*}{REFUGE} & U-Net \cite{ronneberger2015u} & 0.8928 & 0.8529 & 0.8641 & 0.8701 & 0.8444 & 0.86486 & 0.8641 \\
 & U-Net Edge & \textbf{0.9214} & \textbf{0.9161} & \textbf{0.9099} & \textbf{0.9078} & \textbf{0.9236} & \textbf{0.9157} & \textbf{0.9161} \\
\hline
\multirow{2}{*}{Drishti-GS} & U-Net \cite{ronneberger2015u} & \textbf{0.9589} & \textbf{0.9627} & 0.9806 & \textbf{0.9878} & \textbf{0.9922} & \textbf{0.9764} & 0.9806 \\
 & U-Net Edge & 0.9324 & 0.9609 & \textbf{0.9822} & 0.9834 & 0.9918 & 0.9701 & \textbf{0.9822} \\
\hline
\end{tabular}
\end{table*}
The comparison between the edge-integrated U-Net model and the baseline Attention U-Net can provide insights into the impact of edge integration on segmentation performance. Better performance in all the areas of dice scores and Hausdorff distance metrics by the edge-integrated U-Net over the Attention U-Net shows learning edges as targets would vastly make a difference in the learning capabilities of a model.

Considering both the median and mean results when evaluating or interpreting data can provide a more comprehensive understanding of the underlying distribution and performance of models. The median dice score and Hausdorff distance results of all the testing images with and without optic disc and cup edge integration for the optic disc and cup are tabulated in Table \ref{tab:median_opticdisc}. The median results also resonate closely to the mean results. In the case of the REFUGE dataset, there is significant improvement in performance metrics of the U-Net Edge and U-Net++ Edge models compared to the U-Net and U-Net++ models respectively and slight improvement as far as the Attention U-Net Edge model is concerned. On the other hand, there is varied improvement in performance metrics of the U-Net Edge and Attention U-Net Edge models in the case of the Drishti-GS dataset. 

Further, the mean and median optic disc dice score results of the 5-fold cross validation experiment for the U-Net and U-Net Edge models are tabulated in Table \ref{tab:cross_validation}. In the case of the REFUGE dataset, the U-Net Edge model performs consistently better, and its dice score values provide a more accurate and reliable estimate of the model's overall performance and generalizability. In the case of the Drishti-GS dataset, the difference is performance is not vastly different though. 

To understand how the edge-integrated baseline models perform against some state-of-the-art models available in the literature, the authors performed a comparison study between the results of other models and the results of the edge-integrated U-Net model. The comparison results are tabulated in Table \ref{tab:compare_opticdisc}. Only mean dice scores were available for other models in the literature, apart from a few using Jaccard coefficient as well. The use of Hausdorff distance as a performance metric was largely absent in all the literature data available. The authors emphasize the importance of using Hausdorff distance, especially in overlapping segmentation problems since it could provide a completely different context about the model's performance and hope that it is used extensively as a performance metric for future segmentation tasks by researchers worldwide. Also, it has to be noted that the authors are comparing the U-Net Edge model specifically since it is the most baseline of all the models used in the experiments.

The U-Net Edge model performed significantly better compared to other state-of-the-art models in both datasets as far as the mean dice score for the optic cup is concerned. The optic disc mean dice score was comparable, but was not better than other models in the literature. However, it should be noted that incorporating edges improved the performance of the baseline U-Net model significantly, making it comparable to state-of-the-art models found in the literature.

\begin{table*}
\caption{Comparison results against others models in literature}
\label{tab:compare_opticdisc}
\begin{tabularx}{\textwidth}{|l|l|X|X|X|X|}
\hline
 &  & \multicolumn{2}{c|}{Optic Disc} & \multicolumn{2}{c|}{Optic Cup} \\
\hline
Dataset & Model & Dice Score$\uparrow$ & Hausdorff $\downarrow$ & Dice Score$\uparrow$ & Hausdorff$\downarrow$ \\
\hline
\multirow{2}{*}{REFUGE} & M-Net \cite{fu2018joint} & 0.909 & - & 0.844 & - \\
 & Multimap Localization and Conventional U-Net \cite{10322515} & 0.942 & - & 0.843 & - \\
 & Two-stage Mask-RCNN \cite{almubarak2020two} & 0.947 & - & 0.854 & - \\
 & Segtran \cite{li2021medical} & \textbf{0.961} & - & 0.872 & - \\
 & \textbf{U-Net Edge} (Proposed Model) & 0.901 & \textbf{2.826} & \textbf{0.877} & \textbf{2.236} \\
\hline
\multirow{2}{*}{Drishti-GS} & LSACM-SP \cite{zhou2019optic} & 0.955 & - & 0.847 & - \\
 & DeeplabV3+, GAN \cite{wang2019boundary} & 0.961 & - & 0.862 & - \\
 & Multi-feature Based Approach \cite{gao2019accurate} & 0.947 & - & 0.826 & - \\ 
 & Encoder Decoder Net \cite{shankaranarayana2019fully} & \textbf{0.963} & - & 0.848 & - \\
 & \textbf{U-Net Edge} (Proposed Model)& 0.946 & \textbf{12.200} & \textbf{0.910} & \textbf{4.455} \\
\hline
\end{tabularx}
\end{table*}

\subsection{Activation Maps}
\begin{figure}[hbt]
    \centering
    \captionsetup[subfigure]{justification=centering}
    
    \subfloat[Input image]{{\includegraphics[width=0.2\linewidth]{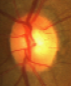} }}
    \hfill 
    \subfloat[Optic disc region]{{\includegraphics[width=0.2\linewidth]{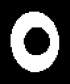} }}
    \hfill 
    \subfloat[U-Net activation map]{{\includegraphics[width=0.2\linewidth]{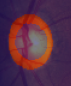} }}
    \hfill 
    \subfloat[U-Net Edge activation map]{{\includegraphics[width=0.2\linewidth]{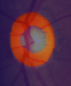} }}
    
    \caption{Sample set of an input image and the corresponding activation maps obtained by U-Net and U-Net Edge models for the optic disc region}
    \label{fig: disc}
\end{figure}

The authors have made use of activation maps to better explain the effect of edge integration on the learning capability of a model. The predictions from the last layer have been used for constructing the activation maps in this study. Figure \ref{fig: disc} (d) and Figure \ref{fig: disc} (c) show the activations obtained by the U-Net and the U-Net Edge (model trained additionally on edge targets) respectively on the input image (Figure \ref{fig: disc}) for the optic disc region. Clearly, more regions of the optic disc are activated in the map generated by U-Net Edge compared to the U-Net model trained without edge targets. 

\begin{figure}[hbt]
    \centering
    \captionsetup[subfigure]{justification=centering}
    
    \subfloat[Input image]{{\includegraphics[width=0.2\linewidth]{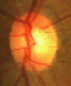} }}
    \hfill 
    \subfloat[Optic cup region]{{\includegraphics[width=0.2\linewidth]{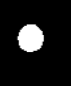} }}
    \hfill 
    \subfloat[U-Net activation map]{{\includegraphics[width=0.2\linewidth]{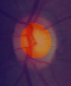} }}
    \hfill 
    \subfloat[U-Net Edge activation map]{{\includegraphics[width=0.2\linewidth]{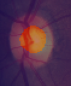} }}
    
    \caption{Sample set of an input image and the corresponding activation maps obtained by U-Net and U-Net Edge models for the optic cup region}
    \label{fig: cup}
\end{figure}

Similarly, Figure \ref{fig: cup} (d) and Figure \ref{fig: cup} (c) show the activations obtained by the U-Net and the U-Net Edge (model trained additionally on edge targets) respectively for the optic cup region. Here, the U-Net model clearly over-segments the optic cup region whereas the U-Net Edge model does not over-segment, as is evident from the activation maps. 

The authors claim from the observations that edge targets provide crucial boundary information that help in distinguishing between different objects or regions in an image, leading to finer segmentation that closely follows the true contours of objects. This could be the primary reason for the reduction of Hausdorff distance scores in models additionally trained on edge targets.

\section{CONCLUSIONS}
\label{con}
Segmenting the optic disc and cup from retinal images is a critical task in diagnosing and managing glaucoma, among other ocular conditions. 
However, this task is fraught with challenges related to image quality, resolution, and the inherent complexity of the eye's anatomy, as well as technical and procedural issues in image processing and analysis.
The presence of blood vessels with varying degrees of tortuosity can obscure the boundaries of the optic disc and cup, complicating their segmentation. Vessels that cross the disc margins can create confusion in automated algorithms, leading to inaccurate segmentation. Images may contain noise and artifacts due to the imaging process, patient movement, or other factors.
Having said that, the authors have tried to address the above difficulties by shifting paradigms and focusing more on the retinal fundus data rather than the architecture of deep learning models. The idea that the optic cup-to-disc ratio (CDR) is one of the most important metrics in the diagnosis of glaucoma and that it solely depends on the boundaries of the optic disc and cup regions has motivated the authors to integrate the regions' edges into the learning process of the model. 

Various popular encoder-decoder models like the U-Net, the Attention U-Net and the U-Net++ have been studied with and without edges as targets along with the optic disc-cup regions. 
The authors have used the REFUGE benchmark dataset and the Drishti-GS dataset to perform the study, and the results are tabulated for the dice and Hausdorff distance metrics. 
In case of the REFUGE dataset, the mean dice score has improved from 0.7425 to 0.8859 while the mean Hausdorff distance has reduced from  6.581 to 3.054 (lesser is better) for optic disc for baseline U-Net model. Additionally, the mean dice score has improved from 0.6970 to 0.8639 while the mean Hausdorff distance has reduced from  5.2340 to 2.6323 (lesser is better) for optic cup for the same model. Similar improvements in results have been observed for the Drishti-GS dataset as well, highlighting the robustness of the approach.

The authors have also studied in detail, the activation maps generated by these models in different combinations and have found edge integration does help prevent over-segmentation and under-segmentation of smaller regions of interest. Additionally, the predicted edges could also help in treatment planning of abnormal images. 

Overall, the workflow proposed can be extended to any 2D image segmentation problem and is expected to improve in cases where the boundaries of individual regions of interest overlap each other.
As a part of future work, the authors would like to experiment with more robust models such as vision transformers. Also, the authors aim to explore other ways of using edges to better train deep learning models.

